\documentclass[utf8]{FrontiersinHarvard}
\usepackage{url,microtype,subcaption}
\usepackage[onehalfspacing]{setspace}
\usepackage{import}
\usepackage[switch]{lineno}
\usepackage[dvipsnames]{xcolor} 
\usepackage{siunitx}
\usepackage{float}
\usepackage{gensymb}
\usepackage{xurl}
\usepackage{graphicx}
\usepackage[normalem]{ulem}
\usepackage{alltt}
\usepackage{hyperref}
\hypersetup{
  colorlinks,
  citecolor=Sepia,
  linkcolor=Sepia,
  urlcolor=Sepia}

\usepackage{tabularx}
\usepackage{appendix}

\def\keyFont{\fontsize{11}{11}\helveticabold}
\def\firstAuthorLast{Kueper, Chari {et~al.}}
\def\Authors{Niklas Kueper\,$^{1,*,\dagger}$, Kartik Chari\,$^{1,*,\dagger}$, Judith Bütefür\,$^{2}$, Julia Habenicht\,$^{2}$, Su Kyoung Kim\,$^{1}$, Tobias Rossol\,$^{1}$, Marc Tabie\,$^{1}$, Frank Kirchner\,$^{1,3}$, and Elsa Andrea Kirchner\,$^{1,2}$}
\def\Address{$^{1}$ Robotics Innovation Centre, German Research Centre for Artificial Intelligence (DFKI), 28359 Bremen, Germany\\
$^{2}$Institute of Medical Technology Systems, University of Duisburg-Essen, 47057 Duisburg, Germany \\
$^{3}$Robotics Lab, University of Bremen, 28359 Bremen, Germany\\
$^{\dagger}$ These authors contributed equally to this work and share first authorship}
\def\corrAuthor{Niklas Kueper}
\def\corrEmail{niklas.kueper@dfki.de}

\begin{document}
\onecolumn
\firstpage{1}

\title {EEG \& EMG dataset for the detection of errors introduced by an active orthosis device} 

\author[\firstAuthorLast ]{\Authors}
\address{}
\correspondance{}
\extraAuth{Kartik Chari\\ kartik.chari@dfki.de}

\maketitle
\begin{abstract}
This paper presents a dataset containing recordings of the electroencephalogram (EEG) and the electromyogram (EMG) from eight subjects who were assisted in moving their right arm by an active orthosis device. The supported movements were elbow joint movements, i.e., flexion and extension of the right arm. While the orthosis was actively moving the subject's arm, some errors were deliberately introduced for a short duration of time. During this time, the orthosis moved in the opposite direction. In this paper, we explain the experimental setup and present some behavioral analyses across all subjects. Additionally, we present an average event-related potential analysis for one subject to offer insights into the data quality and the EEG activity caused by the error introduction. The dataset described herein is openly accessible. The aim of this study was to provide a dataset to the research community, particularly for the development of new methods in the asynchronous detection of erroneous events from the EEG. We are especially interested in the tactile and haptic-mediated recognition of errors, which has not yet been sufficiently investigated in the literature. We hope that the detailed description of the orthosis and the experiment will enable its reproduction and facilitate a systematic investigation of the influencing factors in the detection of erroneous behavior of assistive systems by a large community.

\keyFont{\section{Keywords:} EEG, event-related potential, EMG, orthosis, dataset, robotics, BCI, error detection}
\end{abstract}

\twocolumn
\section{Introduction} \label{section:intro}
Exoskeletons and orthoses are frequently used to assist or enable human movement (see~\citet{kirchner:2022towards}). They are even able to augment classical therapy approaches such as mirror therapy  \citep{Kirchner:2013applicability}. 
The electroencephalogram (EEG) can be used not only to infer the intention to move but also to trigger the assistance provided by an exoskeleton \citep{kirchner:2022towards}. This has been shown to be very important for successful neuro-rehabilitation ~\citep{noda:2012brain,hortal:2015using}. It can also be used to infer the subjective correctness of the behavior of a robot that the human observes or interacts with, as shown in several works such as by~\cite{skim:scirep:2017,skim:icra:2020,chavarriagaErrareMachinaleEst2014}. EEG activity can also be used to enable the teaching of subjective preferences to a prosthesis, as proposed by~\cite{iturrateTeachingBrainmachineInterfaces2015}. Furthermore, misbehavior of an assistive device can be detected using the error-related potentials (\textit{ErrPs}) that occur when the brain observes errors ~\citep{vanSchie2004a}. 

Inferring errors from EEG is challenging because it requires asynchronous classification of relevant patterns in the EEG, which often leads to many false positives due to long interaction times with the system or long task times for the system \citep{omedesAnalysisAsynchronousDetection2015,lopes-diasOnlineAsynchronousDetection2021,SpuelerNiethammer2015,Lopes-Dias_2021}. While most studies focus on how the brain evaluates erroneous behavior using visual stimuli, according to our literature review, there is only one study that uses tactile stimuli to elicit an ErrP \citep{schiattiHumanLoopRobot2018}. Moreover, it has only been shown that rather complex errors in behavior elicit the ErrP or correlated changes in the frequency domain. It has not yet been studied how the brain responds to simpler errors that are quite obvious and may not require extensive evaluation to assess correctness.

Based on the available literature, we believe that there is a gap in the research on error-related activity. Modalities, other than the visual modality, should also be studied more closely during the interaction or as a source of feedback in terms of what activities they induce in the brain. It is also necessary to better understand what complexity in task, response, or interaction errors is required to elicit the ErrP. Furthermore, methods for continuous classification of error-related activity and methods that allow to distinguish between different, partially overlapping EEG activities, should be explored more intensively. In this regard, there is a particular lack of openly accessible data that would allow a larger research community to contribute. Furthermore, the use of robots limits the number of research groups that can conduct research on such problems. For this reason, we recorded a dataset of 8 subjects wearing an active orthosis device that introduces simple errors in its behavior. In addition, details about active elbow orthosis are provided, as it is a fairly simple robotic system that could allow other groups to replicate the work or extend it. Furthermore, the experimental procedure has also been described which would give relevant information about the error complexity, subject instructions, and whether or not subjects should respond explicitly to errors. 

We hope that this first open-access dataset will enable broader research on how assistive technology can be improved by using EEG to provide more natural and individualized support for activities of daily living. Such support is very important for rehabilitation ~\citep{kornhaber:2018resilience}.  

The rest of the paper is structured as follows. Section \ref{section:methods} provides detailed information about the experimental design and methods used to record the dataset. It also describes the data format and the folder structure for a better understanding of the dataset. Furthermore, Section \ref{section:analysis} presents a preliminary quality analysis of the recorded data in the form of response-time analysis and event-related potential analysis. Finally, in Section \ref{section:conclusion}, we provide an overview of this experiment and discuss future possibilities.

\section{Methods and Experimental Design} \label{section:methods}
This section provides information about the experimental design including details about subjects' informed consent, experimental setup \& procedure, methods used for data acquisition, and the formatting of the recorded dataset.

\subsection{Participants} \label{section:participants}
Eight healthy subjects (four male and four female; average age $21.8 \pm 2.4$ years; right-handed; students) voluntarily participated in the study. Some days before the start of experiments, all subjects were invited to the lab for a basic introduction and preliminary testing. This included checking the fit of the used orthosis and measuring the head circumference for determining their EEG cap size. All the subjects gave their written informed consent and were told that they could stop the experiment at any time without consequences. The experiment lasted for $4.9 \pm 0.6$ hours on average and all the subjects received a monetary compensation of 10€ per hour. 

\subsection{Experimental Setup and Procedure} \label{section:exp_setup_proc}
An overview of all the protocols followed throughout the experiment is provided in this section.

\subsubsection{Subject Preparation} \label{section:exp_preparation} 
Before the start of the experiments, the subjects were prepared with a 64-channel EEG system and an eight-channel EMG system (see Section \ref{section:eeg_recording} and \ref{section:emg_recording} for details). Furthermore, each subject was fitted with an active orthosis (see Section \ref{appendix:orthosis}) on their right arm as shown in Figure \ref{fig:exp_setup}-(a), and held a small air-filled ball in their left hand. To trigger support from the orthosis, the subjects were required to express their intention to move by applying a force greater than a start threshold in the movement direction. The required force varied among subjects, depending on their strength. It was ensured that the start thresholds were large enough to prevent unintended starts (refer to Section \ref{appendix:calibration_safety}). After indicating their intention to move, the subjects were instructed to ease their arm muscles as the orthosis took control of the movement and applied adequate torque at the elbow joint. A comprehensive list of the different thresholds for each subject can be found in Table \ref{table:start_thresh}.

\begin{table}
    \renewcommand{\arraystretch}{1.15}
    \caption{Start thresholds for each subject.}
    \label{table:start_thresh}
    \bigskip
    \resizebox{\columnwidth}{!}{
    \begin{tabular}{l|r|r}
        \hline
        \multicolumn{1}{p{1.5cm}|}{\centering \textbf{Subject \\ Code}} & \multicolumn{1}{p{2.9cm}|}{\centering \textbf{Start Threshold \\ (Flexion)}} & \multicolumn{1}{p{2.9cm}}{\centering \textbf{Start Threshold \\ (Extension)}} \\
        \hline \hline
        AQ59D & 1.0 \ \si{\newton\meter} & 1.2 \ \si{\newton\meter} \\
        BY74D & 0.8 \ \si{\newton\meter} & 1.2 \ \si{\newton\meter} \\
        AC17D & 0.8 \ \si{\newton\meter} & 1.2 \ \si{\newton\meter} \\
        AW59D & 0.7 \ \si{\newton\meter} & 1.2 \ \si{\newton\meter} \\
        AY63D & 1.0 \ \si{\newton\meter} & 1.2 \ \si{\newton\meter} \\
        BS34D & 1.2 \ \si{\newton\meter} & 1.4 \ \si{\newton\meter} \\
        AJ05D & 1.0 \ \si{\newton\meter} & 1.2 \ \si{\newton\meter} \\
        AA56D & 1.0 \ \si{\newton\meter} & 1.2 \ \si{\newton\meter} \\
        \hline
    \end{tabular}}
\end{table}

\subsubsection{Experimental Procedure} \label{section:exp_procedure}
\label{exp_procedure}
The subjects' task was to identify errors in the behavior of the orthosis. These errors were deliberately introduced during flexion or extension movements. Here, the term \textit{error} refers to a momentary change in the direction of orthosis movement for a short duration of time (see Table \ref{table:oper_param}). Furthermore, the term \textit{movement trial} will be used to indicate a complete range of flexion \textit{or} extension movement. 

\begin{figure*}
    \centering
    \includegraphics[width=\textwidth]{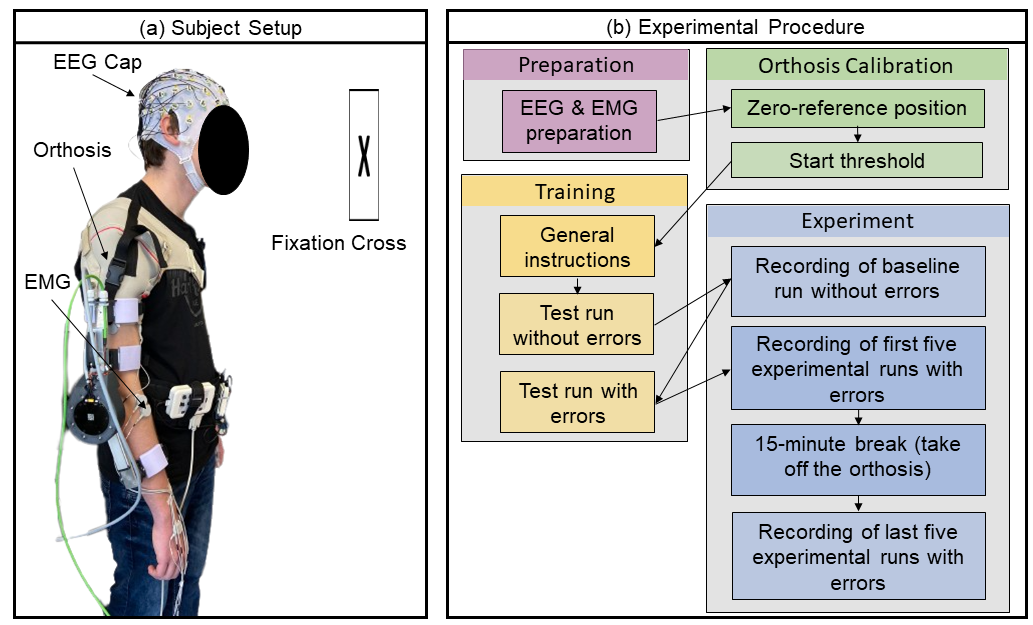}
    \caption{(a) Subject prepared with EEG and EMG electrodes wearing orthosis on their right arm (b) Visualization of the different steps in the experimental procedure.}
    \label{fig:exp_setup}
\end{figure*}

In the first experimental run, subjects were asked to perform 30 movement trials (15 flexions and 15 extensions) with no errors to obtain a baseline. The experimental run began only after the subjects heard a start phrase from the experimenter. This was followed by a training session where the subjects got familiarized with how the errors felt and were instructed to squeeze the ball in their left hand as soon as they felt an error. In each of the following runs, six errors were introduced in a randomized sequence within 30 movement trials. The order of occurrence of errors was varied individually for each run (see Section \ref{section:error_intro} for more details). Overall, as part of the experiment, each subject performed 10 experimental runs with six errors each and one baseline run without any errors.

Before each run, the subjects were reminded to stand still to avoid motion artifacts in the EEG and EMG data. They were also asked to fixate their eyes upon a black cross against the white wall in front of them to avoid eye artifacts in the EEG data. It was also brought to their notice that if, for some reason, they felt an error but forgot to squeeze the ball, they shall just move forward with the run. The subjects were not informed about missed errors during the experiment. At the end of a run, the orthosis motor automatically disabled itself and the subjects were informed about this via a stop phrase. After five runs, a 15-minute break was given during which the subjects could relax and take off the orthosis. A visual summary of the whole experimental procedure is provided in Figure \ref{fig:exp_setup}-(b).

\begin{table}
    \renewcommand{\arraystretch}{1.15}
    \centering
    \caption{Orthosis operating parameters.}
    \label{table:oper_param}
    \bigskip
    \begin{tabular}{l|r}
        \hline
        \multicolumn{1}{p{4.5cm}|}{\centering \textbf{Parameters}} & \multicolumn{1}{p{1.5cm}}{\centering \textbf{Value}} \\
        \hline \hline
        Number of errors & 6 \\
        Duration of errors & 250 ms \\
        Fully Extended position & -10\degree \\
        Fully Flexed position & -90\degree \\
        Maximum deviation & 0.3\degree \\
        Mean error position (Flexion) & -42\degree \\
        Mean error position (Extension) & -58\degree \\
        \hline
    \end{tabular}
\end{table}

\subsubsection{Error Introduction} \label{section:error_intro}
At the start of each experimental run, a distinct list of six random values between 1 and 30 (\textit{number of errors}) was generated. This list, termed as the \textit{error sequence}, followed the following conditions:
\begin{itemize}
    \item Values 1 and 2 must not be included.
    \item There must be a gap of at least two between two consecutive numbers from the list.
    \item The list must be sorted in ascending order.
\end{itemize}
The values in the list corresponded to the trial numbers in which errors were introduced. The Python3 \textit{random} library \citep{python_ref_man} was used to generate the random sequence. 

Furthermore, whenever a movement trial began, the trial number was matched with all numbers from the generated error sequence. If a match was found, an error was deliberately introduced near the \textit{Mean error position} within the movement trial. The error position varied for flexion and extension with a \textit{Maximum deviation} of 0.3\degree \ from the \textit{Mean error position} as mentioned in Table \ref{table:oper_param}. 

In practice, if the orthosis were executing flexion before the introduction of the error, it would transition into extension for the specified duration of error (see Table \ref{table:oper_param}) and then resume flexion until the end of the trial and vice-versa.

\subsection{Data Acquisition} \label{section:data_acquisition}
This section provides detailed information about the methods used for recording the EEG and EMG data. Additionally, it also describes the process of synchronization of these two types of data.

\subsubsection{EEG Recording} \label{section:eeg_recording}
The EEG data were recorded using the 64-channel LiveAmp64 system from Brain Products GmbH\footnote{\url{https://www.brainproducts.com/solutions/liveamp/}} and an ActiCap slim electrode system\footnote{\url{https://www.brainproducts.com/solutions/acticap/}} with an extended 10-20 layout. The reference electrode was placed at FCz and the GND at AFz.

Great efforts were made to record high-quality EEG data and minimize the noise in the data by keeping the impedances of all 64 electrodes below a threshold of \SI{5}{\kilo\ohm}. This impedance check was performed both prior to and after each experimental run. The EEG data were recorded using the \textit{Recorder} software\footnote{\url{https://www.brainproducts.com/downloads/recorder/}} (version 1.25.0001) from Brain Products GmbH. The sampling rate was \SI{500}{\hertz} and the measurement system used hardware filters that limited the bandwidth of the data to a passband of \SI{0.0}{\hertz} - \SI{131.0}{\hertz}.

\subsubsection{EMG Recording} \label{section:emg_recording}
To record bipolar EMG data, the ANT mini eego amplifier\footnote{\url{https://www.ant-neuro.com/products/eego_8}} was used. The EMG data were recorded with a sampling rate of 1000 Hz using an adapted eego SDK\footnote{\url{https://gitlab.com/smeeze/eego-sdk-pybind11/-/tree/0ace9b329b7cf5f6d1da5d387d0f2a5c07e87ee7}} for Python. Eight channels were used, each measuring the muscle activity of the following muscles on both the arms:
\begin{itemize}
    \item M. biceps brachii
    \item M. triceps brachii lateral
    \item M. triceps brachii long head
    \item M. flexor digitorum superficialis
\end{itemize}

Before placing the electrodes, the skin was prepared with Isopropyl alcohol (70\% V/V). The electrodes were placed on the muscle belly in accordance with the SENIAM guidelines \citep{Hermens2000}.

\subsubsection{Synchronization of EMG \& EEG Data} \label{section:data_sync}
To enable synchronization of EEG and EMG data for offline analysis, an Arduino Nano board and the Sensor \& Trigger extension from the EEG system were used to mark start and end time points of the EMG data recordings within the EEG data. The EEG system was used as the main device to enable an alignment of both data streams with respect to each other. Despite the EEG data recordings starting before the EMG, the marked events serve as reference points to align both data streams. With this approach, an average time difference below \SI{8.5}{\milli\second} between both data streams was achieved after evaluating the synchronicity for all recorded data sets. This result was arrived at by comparing the amount of recorded EMG data against the marked events recorded by the EEG system. Please refer to section \ref{section:recorded_events} for the specification of the marked events in the EEG data.

\subsection{Dataset and Format} \label{section:dataset_format}
This section provides a description of the data format, along with detailed information about the dataset and the recorded events.

\subsubsection{Data Format} \label{section:data_format}
The recorded EEG data follows the BrainVision Core Data Format 1.0, consisting of a binary data file (\textit{.eeg}), a header file (\textit{.vhdr}), and a marker file (\textit{.vmrk})\footnote{\url{https://www.brainproducts.com/support-resources/brainvision-core-data-format-1-0/}}. For ease of use, the data can be exported into the widely adopted BIDS format as described in \cite{gorgolewski2016brain}. Furthermore, for data analysis, processing and classification, two popular options are available - MNE (Python)\footnote{\url{https://mne.tools/stable/reading_raw_data.html}} and EEGLAB (MATLAB)\footnote{\url{https://sccn.ucsd.edu/eeglab/index.php}}. In contrast, the EMG data is stored in the \textit{.txt} format, where each column represents a separate EMG channel.

\subsubsection{Dataset Description} \label{section:dataset_description}
In this section, the dataset's folder structure is explained along with the convention used for naming the files.
\paragraph{Folder Structure} \label{para:folder_structure}
This section describes the hierarchical folder structure of the recorded dataset. At the highest level, there are three folders, namely \textit{EEG}, \textit{EMG}, and \textit{Metadata}. The \textit{Metadata} folder contains a \textit{.txt} file for each subject, segregated by a unique code, which consists of meta-information about the subject as well as the measurement sets. In addition to these files, there is also a \textit{short\_description.txt} file with some general information about the whole study.

Furthermore, within each of the modality folders (\textit{EEG} or \textit{EMG}), there is an additional level of folders segregated by subject codes. Inside the \textit{EEG} folder, each subject sub-folder is further divided into two sub-folders viz. \textit{data} and \textit{imp}. The \textit{data} folder consists of the actual measurement files as described in section \ref{section:data_format}. In total, there is one baseline set without any errors stored inside another sub-folder named \textit{baseline\_without\_error} and 10 sets with deliberate errors introduced. Each header file (\textit{.vhdr}) also contains the impedance values of every electrode before the set. Conversely, each header file inside the \textit{imp} folder contains impedance values after each set. All in all, all impedance values, before and after the set, are available within the header files (\textit{.vhdr}). It has to be noted that, for some subjects, an additional set was recorded for safety purposes and included in this dataset under a sub-folder named \textit{additional sets}. For more detailed information, please refer to the Metadata readme files included within the dataset.

\paragraph{Naming Convention for Data Files}
A consistent naming convention was followed for all our files dividing the filename into five parts. The first part was the date of acquisition in \textit{yyyymmdd} format (e.g.\ 20230424), followed by the subject code (e.g.\ AC17D). The third part included the experiment identifier, in this case, \textit{orthosisErrorIjcai}, followed by \textit{multi} indicating that subjects could hear and see (multiple modalities) during the experiments. For baseline runs, the suffix \textit{baseline\_set} plus the set number (e.g \textit{1} or \textit{2}) was added while for experimental runs with errors, only the run number was appended at the end (e.g.\ \textit{set5}). As an example, a filename would look like \textit{20230424\_AC17D\_orthosisErrorIjcai\_multi\_set1.txt}. It is important to note that, the term \textit{set} was used to represent the data files associated with the corresponding experimental run.

\subsubsection{Recorded Events} \label{section:recorded_events}
To keep track of all the events occurring during the set, these events were recorded and stored in marker files (\textit{.vmrk}). The marker files are located within the \textit{data} folder of each subject under \textit{EEG} (see section \ref{section:dataset_description} for data structure). The first recorded event (after the start of a set) was named \textit{S1} and it marked the start of the EMG recording for synchronization purposes (see section \ref{section:data_sync} for detailed information). The event \textit{S1} also occurred at the end of the EMG measurement. The next recorded event was \textit{S64} which marked the start of flexion movement. Similarly, the start of an extension movement was marked by the event \textit{S32}. In order to mark a trial without errors, the event \textit{S48} was added around the \textit{Mean error position} as mentioned in Table \ref{table:oper_param}. The event \textit{S96} occurred as soon as an error was introduced in the trial. Additionally, if the subject squeezed the ball, the recorded event \textit{S80} was written to the marker file.

\section{Analysis of Data Quality} \label{section:analysis}
In the following, we performed some basic analysis on the recorded data to give proof of data quality and to briefly describe evoked EEG activity as well as muscle activity. 
Furthermore, to ensure data quality, invalid measurement sets were excluded from the data repository. The excluded sets of all subjects are listed below in the form of \emph{subject code}, \emph{set number} followed by the reason for excluding the set. The excluded sets are as follows: 
\begin{itemize}
    \item \emph{AA56D, set8}: orthosis shutdown by max. current limit.
    \item \emph{AC17D, baseline\_set1}: too much noise in EEG.
    \item \emph{AJ05D, baseline\_set1}: error in event recording.
    \item \emph{AJ05D, set8}: too much noise in the EEG.
    \item \emph{AJ05D, set9}: too much noise in the EEG.
    \item \emph{AQ59D, set1}: orthosis did not start.
    \item \emph{AW59D, set1}: too much noise in the EEG.
    \item \emph{AY63D, baseline\_set1}: subject played with the air-filled ball during experiment.
\end{itemize}
Each of the rejected sets was excluded and supplemented by an additional measurement set (as mentioned in Section \ref{para:folder_structure}). 
Although the data quality was kept as high as possible, we observed a 50 Hz noise for some measurement sets and EEG channels, introduced by the orthosis. The 50 Hz noise was also observed in the EMG data. 

\subsection{Behavioral analysis}
For response-time analysis, we analyzed the response times for the incorrect events (error events). 
As mentioned above, the subjects were instructed to squeeze an air-filled ball after recognizing an error. The time between the error event (S96, true label) and response to the event (S80) was calculated for all events. 

According to the experimental design, we expected a total of 480 responses to error events (6 error events $\times$ 10 datasets $\times$ 8 subjects = 480 error events). 
However, we found 9 false negative cases (i.e., the ball was not squeezed, even after an error event occurred) and 5 false positive cases (i.e., the ball was squeezed, even when the error event did not occur). 
Hence, a total of 471 error event-response pairs (480 error events - 9 false negatives) were used to compute response time. 
We obtained a median value of \SI{0.72}{\second} over 471 error event-response pairs. 

We also performed two additional analyses. 
First, we calculated the response time averaged over all 10 sets for each subject (see, Table \ref{table:rt_results}-(A)). 
We also calculated the response time averaged over all 8 subjects for each set (see, Table \ref{table:rt_results}-(B)).

\begin{table}
    \renewcommand{\arraystretch}{1.15}
    \centering
    \caption{Results of response-time (RT) analysis. (A) Median RT for each subject over 10 datasets. (B) Median RT for each set over 8 subjects. $\mu\pm\sigma$ : mean $\pm$ standard deviation.}
    \label{table:rt_results}
    \bigskip
    \begin{tabular}{c|c||c|c}
        \hline
        \multicolumn{2}{p{3cm}||}{\centering \textbf{(A) Analysis 1}} & \multicolumn{2}{p{3cm}}{\centering \textbf{(B) Analysis 2}}\\
        \hline
        \multicolumn{1}{p{1.3cm}|}{\centering \textbf{subject}} & \multicolumn{1}{p{1.4cm}||}{\centering\textbf{RT}} & \multicolumn{1}{p{1.2cm}|}{\centering \textbf{dataset}} & \multicolumn{1}{p{1.4cm}}{\centering\textbf{RT}}\\
        \hline \hline
        AQ59D & 0.67 \si{\second}& set 1 & 0.72 \si{\second}\\
        BY74D & 0.61 \si{\second}& set 2 & 0.73 \si{\second}\\
        AC17D & 0.83 \si{\second}& set 3 & 0.70 \si{\second}\\
        AW59D & 0.72 \si{\second}& set 4 & 0.71 \si{\second}\\
        AY64D & 0.67 \si{\second}& set 5 & 0.70 \si{\second}\\
	BS34D & 0.68 \si{\second}& set 7 & 0.74 \si{\second}\\
        AJ05D & 0.91 \si{\second}& set 6 & 0.80 \si{\second}\\
	AA56D & 0.89 \si{\second}& set 8 & 0.76 \si{\second}\\
             &     & set 9 & 0.68 \si{\second}\\
             &     & set 10 & 0.74 \si{\second}\\
        \hline \hline
        $\mu\pm\sigma$ & 0.75$\pm$0.11 & 
        $\mu\pm\sigma$ & 0.73$\pm$0.03 \\
        \hline
    \end{tabular}
\end{table}

\subsection{Event Related Potentials (ERP) Analysis}

For ERP analysis, we analyzed the EEG data using EEGLAB\footnote{\url{https://sccn.ucsd.edu/eeglab/index.php}}. 
We preprocessed the data as follows.
The raw EEGs were downsampled to \SI{250}{\hertz}, re-referenced to an average reference, and filtered between \SI{0.1}{\hertz} and \SI{15}{\hertz}. 
The \emph{FCz} channel, used as a reference in the EEG recording, was recalculated as an EEG channel for ERP analysis. 
After preprocessing, the artifacts were rejected by visual inspections which means that only clean EEG data were used for EEG segmentation. 
The artifact-free EEG data were segmented into epochs from \SI{0.1}{\second} to \SI{1}{\second} after each event type (correct/incorrect). 
Epochs were averaged within each event type with a baseline correction (\SI{-0.1}{\second} until stimulus onset).

\begin{figure*}
    \centering
    \includegraphics[width=\textwidth]{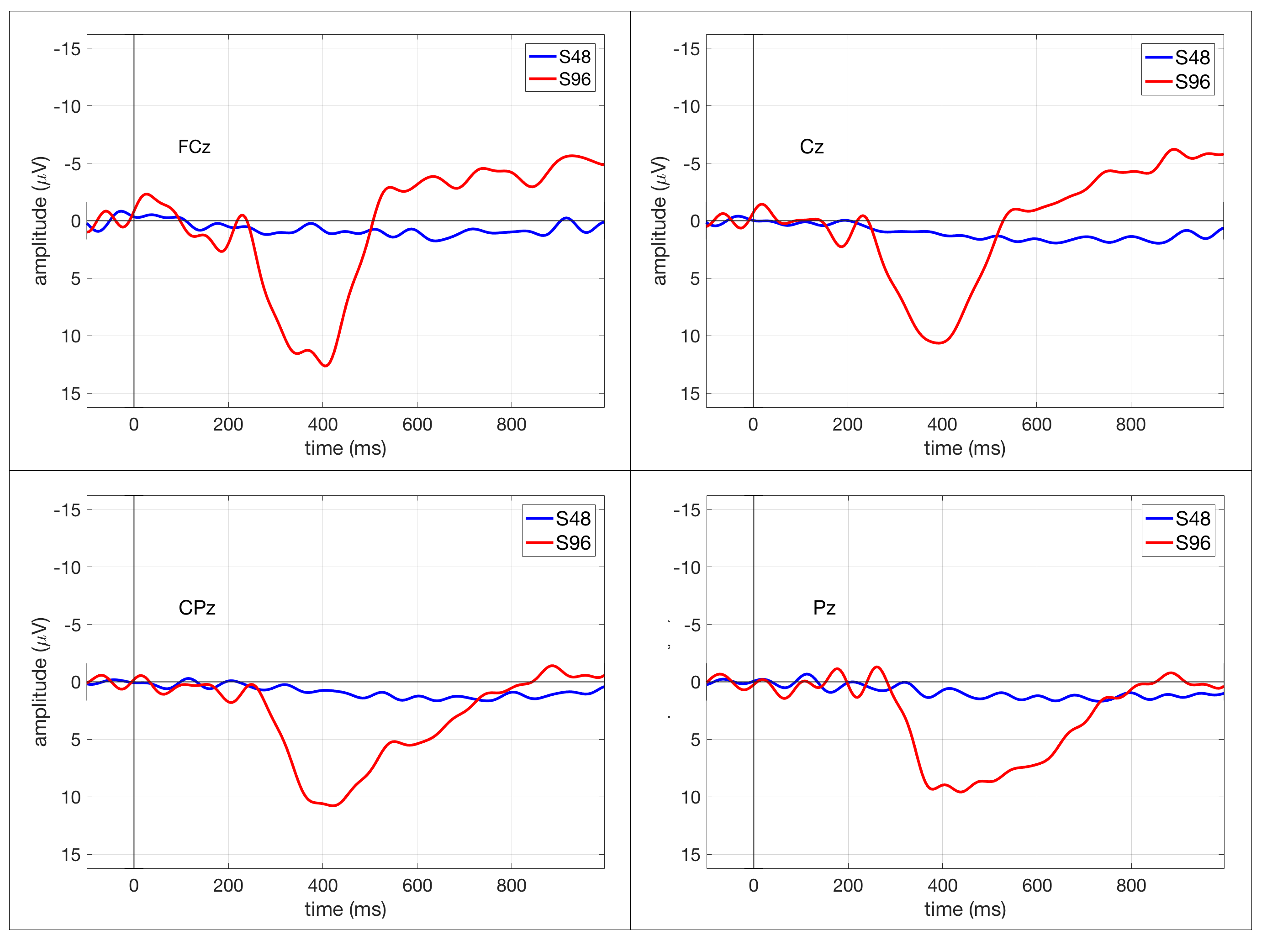}
    \caption{ERPs averaged over all epochs (trials) within each event type: correct event (S48) and incorrect event (S96) for Subject AQ59D.}
    \label{fig:erp_results}
\end{figure*}

For averaging epochs, we only used the error events with correct responses i.e., true positive cases (the ball was squeezed when error events occurred). 
Figure \ref{fig:erp_results} shows the ERPs averaged over all epochs for each event type (S48: correct event, S96: incorrect event) for Subject AQ59D. The ERP morphology, i.e., the shape and distribution on the scalp suggest that the introduction of errors elicits a P300 component, more specifically a P3b \citep{PolichJohn2007UPAi} component. This may be elicited by infrequently occurring odd events to which subjects respond, i.e., task-relevant events \citep{10.3389/fnhum.2016.00291}. In the case of this subject, we could not observe an error-related potential, which is usually evoked by the recognition of errors. 

\section{Conclusion} \label{section:conclusion}
We presented and described an open-access dataset of EEG and EMG data obtained from eight subjects who were assisted by an active orthosis device in moving their right arm. Behavioral analysis showed that the errors were very well recognized by the subjects. Introduced errors were a momentary change in the direction of movement of the orthosis for a short amount of time. The errors were simple and easy to detect. The appearance of an average event-related potential (ErrP) in the form of \emph{P3b} indicates that the subject recognized the erroneous events as odd events. One reason that the error introduction did not elicit an error-related potential in the EEG could be due to the simplicity of the error. On the other hand, the very dominant \emph{P3b} could also overlay the ErrP. These conclusions are very preliminary due to the analysis of only one subject. We hope that the dataset provided and the detailed information about the experimental setup would allow its replication. This would enable the research community to systematically investigate the relationship between odd-event detection and erroneous event evaluation, evoked in the brain. A better understanding of this relationship would help to develop future approaches that could allow automatic adaptation of an assistive device to a subject's subjective needs.

\section*{Data Availability Statement}
The dataset recorded in this study can be found online on Zenodo\footnote{Training data: \url{https://doi.org/10.5281/zenodo.7951044}} \footnote{Test data: \url{https://doi.org/10.5281/zenodo.7966275}}. 

\section*{Ethics Statement}
The studies involving human participants were approved by the local Ethical Committee of the University of Duisburg-Essen, Germany. The participants provided their written informed consent to participate in this study.

\section*{Author Contribution}
EK and SK: designed the study. FK: provided feedback to the study design. NK, JH, JB, and KC: prepared the subjects and conducted the experiments. KC and NK: developed the software for operating the experimental setup. TR: responsible for the mechatronic design of the active orthosis device. TR and MT: set up the low-level control and interface of the elbow motor. SK and NK: analyzed the data quality (response-time analysis and ERP analysis) and wrote the results. NK and JB: managed the dataset. NK, KC, JB, JH, and TR: drafted the manuscript. EK: wrote the abstract, introduction and conclusion. EK, FK and KC: provided critical revisions. KC: handled the overall formatting of the manuscript. EK: supervised the study. All authors contributed to the report and approved the submitted version.

\section*{Funding}
This work was funded by the German Federal Ministry of Education and Research (BMBF) within the project M-RoCK (Grant number: 01IW21002) and by the Institute of Medical Technology Systems at the University of Duisburg-Essen. 

\section*{Conflict of Interest Statement}
The authors declare that the research was conducted in the absence of any commercial or financial relationships that could be construed as a potential conflict of interest.

\section*{Acknowledgement}
We would like to express our sincere gratitude to all the subjects who participated in the study. Furthermore, we would like to thank the students from the University of Duisburg-Essen, namely Variscia Permata Putri, Patrick Fugmann, and Sophia Theisinger who conducted preliminary experiments and provided valuable insights that helped us greatly with this study. Additionally, we would like to acknowledge Jan-Philipp Brettschneider for his work on the mechanical design of orthosis while interning at DFKI.

\appendix

\renewcommand{\thesubsection}{\Alph{subsection}}

\section*{\centering Appendix}
\begin{appendices}

\subsection{Active Orthosis Device} \label{appendix:orthosis}
\begin{figure*}
    \includegraphics[width=0.9\textwidth]{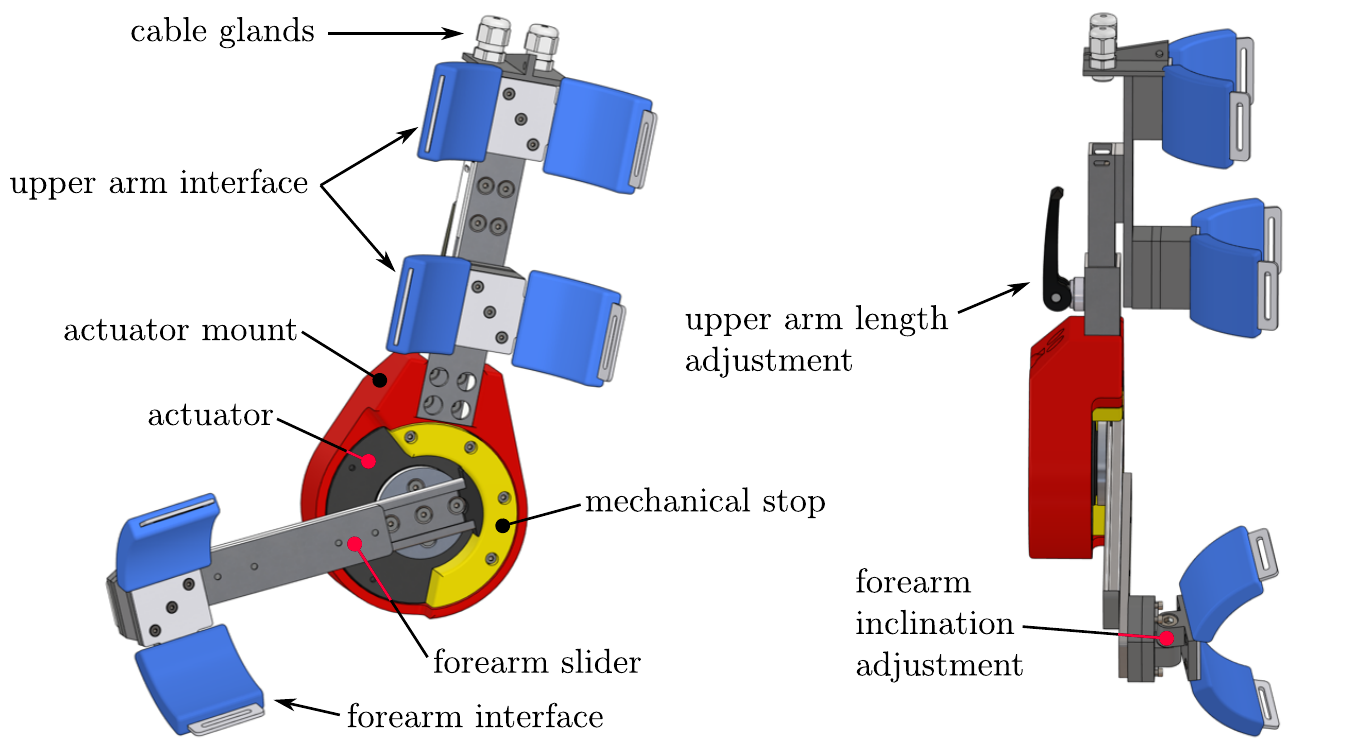}
    \caption{Computer Aided Design (CAD) model side and front view of the active orthosis device. The actuator mount and mechanical stop are colored for better visibility. The connection to the shoulder orthosis can be seen in Figure \ref{fig:exp_setup}-(a).}
    \label{fig:mech_design}
\end{figure*}

The orthosis device (Figure \ref{fig:mech_design}) was mainly built from off-the-shelf components and in-house manufactured parts (e.g. 3D printed or water jet cut) resulting in a simple and cost-effective design. The orthosis consists of an upper arm and a forearm structure connected and driven by an actuator. To compensate for the weight of the device and thereby enhance wearer comfort, the upper arm structure is attached to a strap that is sewn on a shoulder orthosis (see Figure \ref{fig:exp_setup}-(a)).

To transmit forces to the wearer's arm, the elbow orthosis connects to the human body via an upper arm and a forearm interface. The cuffs consist of silicone pads cast on bent, water-jet-cut aluminum sheet parts. The sheet parts transmit the interaction forces between the orthosis and the wearer and can be deformed elastically to an individual arm shape. The silicone pads not only provide soft contact points to the human tissue but also friction between the arm and interfaces to keep the orthosis in place. 

To accommodate different wearers, both upper arm and forearm structure lengths can be adjusted. While the upper arm length is fixed after adjustment, the forearm interface is attached to a slider. This prismatic joint adds a passive degree of freedom to the system that not only adjusts the forearm length but also compensates for misalignments between the axes of rotation of the elbow and orthosis. For further individual adjustment, the forearm interface can be inclined. 

The actuator is a T-MOTOR AK80-6 with a rated torque of \SI{6}{\newton\meter} (\SI{12}{\newton\meter} peak). This drive belongs to the category of so-called quasi-direct-drives or proprioceptive actuators (\cite{Wensing2017}). These comprise a high torque density motor combined with a low gear ratio transmission. Such actuators feature high intrinsic back-drivability and simultaneously allow for accurate high-bandwidth force control. The quasi-direct-drive actuation paradigm has been introduced in rehabilitation robotics throughout the last years, e.g., in \cite{Lv2018, Yu2020}. For communication with the AK80-6, we used a USB to CAN interface and the driver software of \cite{Vyas2023}. We operated the drive in position control mode using the onboard low-level controller. In addition to the software limits, a mechanical stop constrains the range of motion of the orthosis device from the hard limit (upper arm and forearm structure in parallel) to a flexion of \SI{-130}{\degree}, for safety reasons. 

The orthosis was powered by a laboratory power supply set to a voltage of \SI{24}{\volt} and a current of up to \SI{3}{\ampere}. An emergency stop was connected in series with the actuator in case of malfunctions, and a \SI{2400}{\micro\farad}, \SI{100}{\volt} rated capacitor was also added in parallel with the actuator for transient voltage suppression. Cable glands were used as strain reliefs to protect the data and power cable connectors from damage.

\subsection{Response Listener and Event Trigger Board} \label{appendix:button_process}
Whenever the subject felt an error in the orthosis device, they would press the air-filled ball in their left hand. The state of this ball was recorded by the \textit{Response Listener}, a microcontroller that continuously read 16-bit analog inputs and transmitted the pin states, through a serial connection, to the Python script. This script then converted the pin states into single-byte values and sent them serially to the \textit{Event Trigger Board} which is an ATmega328-based microcontroller board (Arduino Nano\footnote{\url{https://store.arduino.cc/products/arduino-nano}}). 

Additionally, other events such as the error introduction, start of flexion or extension, and no-error movement trial were also sent as unique single-byte arrays to the \textit{Event Trigger Board}. This board read the byte values serially and mapped them into transistor-transistor logic (TTL) output signals. These signals were eventually recorded into the electroencephalogram (EEG) marker files, as described in the \textit{Recorded Events} section of the paper. 

\subsection{Calibration Sequence and Safety Measures} \label{appendix:calibration_safety}
\begin{figure*}
    \centering
    \includegraphics[width=0.8\textwidth]{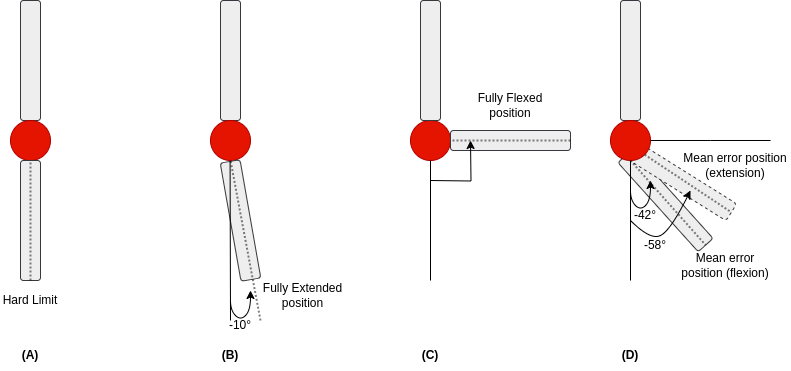}
    \caption{Different operating positions of the orthosis. \textbf{(A)} Zero-reference position for the experimental run. \textbf{(B)} Comfortable start position and extension limit during movement trial. \textbf{(C)} Flexion limit during movement trial. \textbf{(D)} Positions around which errors were introduced.}
    \label{fig:orthosis_oper}
\end{figure*}

For calibrating the zero-reference position of the orthosis, the subjects were asked to fully extend their arms to the hardware limit of the elbow motor. After this calibration, the orthosis would move to a more comfortable start position which was termed the \textit{Fully Extended position} (see Figure \ref{fig:orthosis_oper}). As a safety measure, the range of motion of the orthosis was restricted between the \textit{Fully Extended position} and \textit{Fully Flexed position} (both inclusive). On reaching the range limits, the elbow motor stopped automatically and held the position until the subjects crossed the start threshold. Moreover, another layer of safety was added by restricting the maximum supply current and forcing motor torque limits.

Furthermore, the start thresholds for flexion and extension were calibrated individually. Initially, the threshold values were set to \SI{1.0}{\newton\meter} for flexion and \SI{1.2}{\newton\meter} for extension. To determine the appropriate threshold value for flexion, multiple flexion movement trials were performed by the subjects until they arrived at a force that felt neither too weak nor too strong. In contrast, to establish the optimal start threshold for each subject for extension, they were asked to rest their arms in the \textit{Fully Flexed position} and the lowest value at which no unwanted extensions occurred was selected. 

\bibliographystyle{agsm}
\bibliography{references}

\end{appendices}

\end{document}